\documentclass[10pt,english]{iopart}
\usepackage{iopams}
\usepackage[T1]{fontenc}
\usepackage[latin9]{inputenc}
\usepackage{float}
\usepackage{graphicx}
\usepackage{esint}

\usepackage{xcolor}
\definecolor{darkblue}{rgb}{0,0,0.5}
\usepackage{transparent}
\newcommand{\text}{\rm}

\usepackage{dblfloatfix}
\usepackage{fixltx2e}

\usepackage{babel}

\begin{document}

\title[]{Quantum Feedback Cooling of a Mechanical Oscillator Using Variational
Measurements: Tweaking Heisenberg's Microscope}

\author{Hojat Habibi,$^{1,2}$ Emil Zeuthen,$^2$ Majid Ghanaatshoar,$^1$ and Klemens Hammerer$^2$}
\address{$^1$Laser and Plasma Research Institute, Shahid Beheshti University,
G. C., Evin 1983969411, Tehran, Iran\\
$^2$Institute for Theoretical Physics \& Institute for Gravitational
Physics (Albert-Einstein-Institute), Leibniz Universit\"{a}t Hannover,
Callinstra{\ss}e 38, 30167 Hannover, Germany}

\begin{abstract}
We revisit the problem of preparing a mechanical oscillator in the
vicinity of its quantum-mechanical ground state by means of feedback
cooling based on continuous optical detection of the oscillator position.
In the parameter regime relevant to ground state cooling, the optical back-action and imprecision
noise set the bottleneck of achievable cooling and must be carefully
balanced. This can be achieved by adapting the phase of the local oscillator in the homodyne detection realizing a so-called variational measurement. The trade-off between accurate position measurement and minimal disturbance
can be understood in terms of Heisenberg's microscope and becomes
particularly relevant when the measurement and feedback processes
happen to be fast within the quantum coherence time of the system
to be cooled. This corresponds to the regime of large quantum cooperativity
$C_{\text{q}}\gtrsim1$, which was achieved in recent experiments
on feedback cooling. Our method provides a simple
path to further pushing the limits of current state-of-the-art experiments
in quantum optomechanics.
\end{abstract}
\ioptwocol

\section{Introduction}

\begin{figure}[b]
\begin{centering}
\includegraphics[width=1\columnwidth]{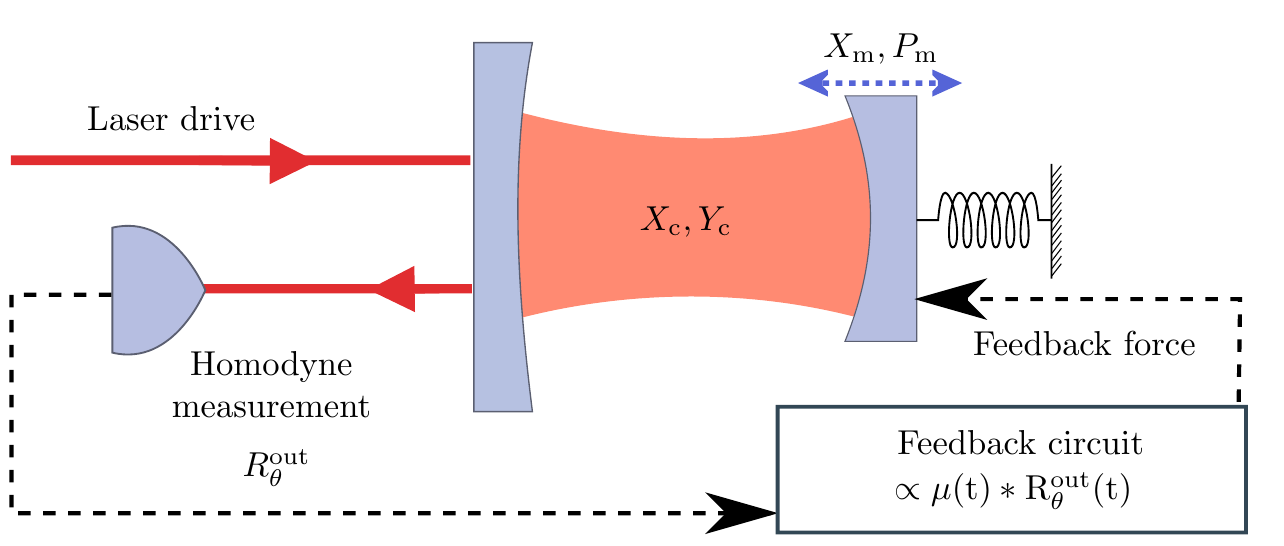}

\caption{Optomechanical system with feedback. The mechanical motion of
the end-mirror in a Fabry-P\'{e}rot cavity mirror of position $X_{\text{m}}$
and momentum $P_{\text{m}}$ is radiation-pressure coupled to a single
cavity mode described by amplitude and phase quadratures $X_{\text{c}},Y_{\text{c}}$.
The light serves as a meter field, which is continually read out from the cavity, and a particular quadrature $R_{\theta}^{\text{out}}$ is obtained by homodyne measurement at local oscillator phase $\theta$. Based on the measurement record, that is determined by a spectral gain function $\mu(\text{t})$ programmed into the feedback circuit, a feedback force is applied to the mechanically compliant mirror ideally steering it into its ground state.  \label{fig:Optomechanical-simple}}
\end{centering}
\end{figure}

The task of exerting quantum-level control over the motion of mechanically
compliant elements has become a central challenge in several fields
of physics ranging from quantum-limited measurement of the motion
of kilogram-scale mirrors in laser-interferometric gravitational wave
detectors \cite{Danilishin2012,Chen2013} to experiments with nano-
and micromechanical oscillators in optomechanics \cite{Aspelmeyer2010,Aspelmeyer2013}.
A paradigmatic example of such a system is an optical cavity mode
coupling via radiation pressure to a mechanical mode whose motion
modulates the optical resonance frequency (see Fig.~\ref{fig:Optomechanical-simple}). Current experiments in this direction involving meso- and microscopic
oscillators include implementations of state-transfer \cite{OConnell2010,Palomaki2013},
frequency conversion \cite{Bochmann2013,Bagci2014,Andrews2014}, impulse
force measurement \cite{HosseiniGuccioneSlatyerEtAl2014}, dynamical
back-action cooling to near the quantum mechanical ground state \cite{Rocheleau2010,Teufel2011,Chan2011,Verhagen2012},
ponderomotive squeezing of light \cite{Brooks2012,Safavi-Naeini2013,PurdyYuPetersonEtAl2013},
and the generation of nonclassical \cite{Riedinger2016}, squeezed
\cite{Wollman2015} and entangled states \cite{Palomaki2013a} of
mechanical oscillators. Of these, the capability to perform ground-state
cooling is the most straightforward benchmark of a quantum-enabled
system, and it is this task that we will consider in the present paper.

Generally, two main approaches to optomechanical cooling have been
considered in the literature, being of respectively passive and active
nature: Dynamical back-action cooling and feedback cooling \cite{Genes2008}.
The former relies on overcoupling the mechanical mode to a ``cold''
reservoir (e.g. optical vacuum) to which it will equilibrate. Meanwhile,
feedback cooling works by continuously measuring the oscillator motion
and, conditioned on the result, applying a force to the oscillator
by some auxiliary means. In either scheme the cooling can be understood as an attempt to map the state of the quiet meter field onto the mechanical mode faster than
the thermal decoherence rate of the latter. Roughly speaking, these
approaches are preferred in the resolved- and unresolved-sideband
regimes, respectively \cite{Genes2008}. Whereas dynamical back-action
cooling to the vicinity of the ground state has already been successfully
demonstrated in numerous experiments \cite{Rocheleau2010,Teufel2011,Chan2011,Verhagen2012},
it is only recently that active feedback cooling has started to approach
the quantum regime \cite{WilsonSudhirPiroEtAl2015,SudhirWilsonSchillingEtAl2016}.
Reflecting this circumstance, the theory of quantum feedback cooling
has not been explored to the same extent as that of the passive approach,
and it is here the present work seeks to contribute.

\begin{figure}[t]
\begin{centering}
\includegraphics[width=1\columnwidth]{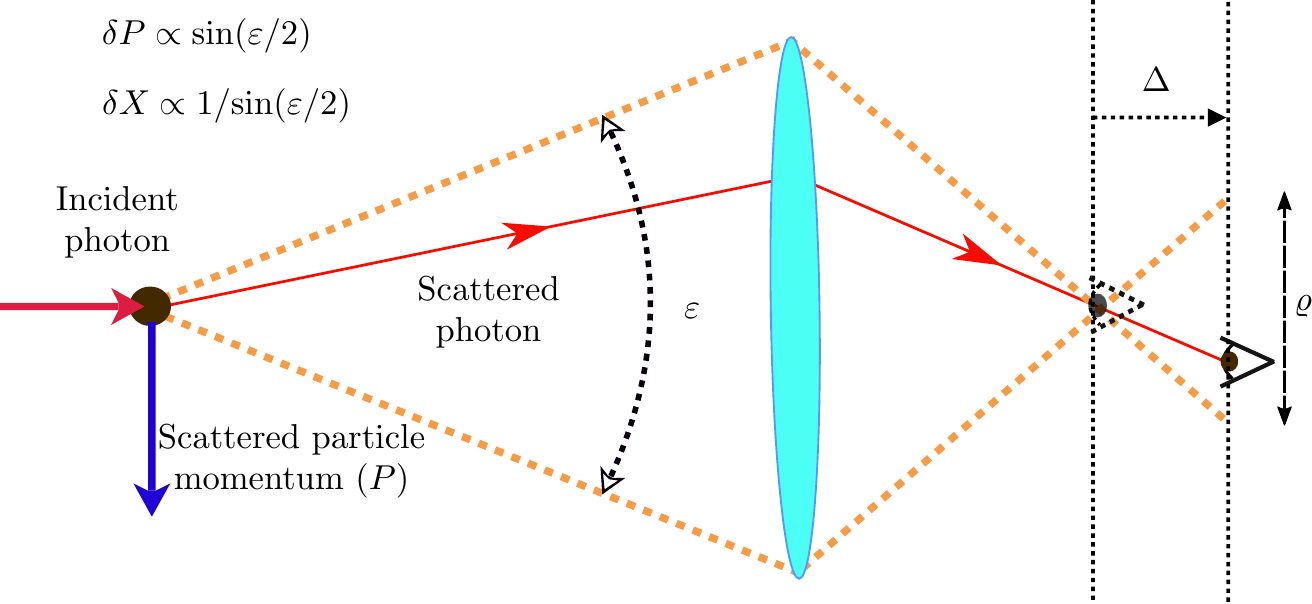}

\caption{Heisenberg's
microscope modified by using a variational measurement that simultaneously
obtains information about position and momentum: A particle (black
dot) is confined to the focal plane of the lens (turquoise ellipse).
A photon incident from the left scatters off the particle giving it
a momentum kick. Photons within the angle $\varepsilon$ are collected
by the lens and refocused at the image plane (eye, dashed). In this
configuration, the image resolution is set by the diffraction limit
of light, $\delta X\propto1/\sin(\varepsilon/2)$, whereas the uncertainty
in momentum introduced by the scattering is $\delta P\propto\sin(\varepsilon/2)$.
Moving the observation plane a distance $\Delta$ out of focus decreases
the position resolution ($\propto1/\varrho$) but allow the observer (eye, solid) to partially
resolve where in the plane the scattered photon arrived.\label{fig:Heisenberg microscope-simple}}
\end{centering}
\end{figure}

In feedback cooling schemes, a balance must be sought between the level of precision at which  the system of interest is monitored and the level of disturbance introduced to it. In the context of optomechanics, the measurement imprecision is set by the vacuum fluctuations of the measured field quadrature while measurement back action is due to vacuum fluctuations in the amplitude quadrature. The trade-off between measurement error and disturbance is a familiar theme within the realm of quantum mechanics which is well illustrated by the famous Gedankenexperiment on the Heisenberg microscope (Fig.~\ref{fig:Heisenberg microscope-simple}) and is expressed quantitatively
in quantum measurement theory in terms of the standard quantum limit
for continuous measurements \cite{Danilishin2012} and error-disturbance
relations \cite{Busch2014}. In the context of feedback cooling, it
will be vital to choose the right trade-off between measurement precision
and disturbance of the system in order to minimize the effective mechanical
temperature. This issue becomes relevant when the measurement and
feedback processes happen to be fast compared to the quantum coherence
time of the system to be cooled as demonstrated in the recent experiment
by Wilson et al.~\cite{WilsonSudhirPiroEtAl2015,SudhirWilsonSchillingEtAl2016}.
We note that this is equivalent to the regime of so-called strong
optomechanical cooperativity which has been achieved, albeit in other
contexts, in several of the experiments cited above. In view of this
recent experimental progress we will address here one particular easily-implementable
method to balance error and disturbance, namely the adjustment of
the local oscillator phase in the homodyne detection of light. This
method has been suggested first in the context of gravitational-wave
detectors by Vyatchanin \cite{Vyatchanin1993} and Kimble \cite{Kimble2001}.
It has also been suggested in our specific context of quantum feedback
cooling in Refs.~\cite{Genes200933,Hofer2015}, where it was shown theoretically
to give an advantage over the conventional phase choice for the local
oscillator. Here, we will provide a systematic optimization of the scheme, which has not been given so far and, moreover, we put the method in the appropriate
conceptual framework of measurement error and disturbance.

In the next section we will introduce a model system for feedback
cooling in optomechanics after which we will present its solution in Sec.~\ref{sec:Mech-resp}.
Before proceeding with the rigorous analysis, we provide in Sec.~\ref{sec:Heisenberg-microscope}
an intuitive explanation of the error-disturbance balance in feedback
cooling and the role of using variational measurements. Then in Secs.~\ref{sec:Steady-state-occupation}
and \ref{sec:Optimized-cooling} we calculate and minimize the effective
mechanical temperature in the presence of variational measurements.
Finally, we conclude and give an outlook in Sec.~\ref{sec:Conclusion}.

\section{Optomechanical equations of motion with feedback\label{sec:Model}}

We will now present the model system to be analyzed. As schematically
shown in Fig.~\ref{fig:Optomechanical-simple}, we study the standard optomechanical
setup consisting of a Fabry-Pérot cavity with a resonating mirror,
but the treatment is applicable to other types of optomechanical setups.
The optical output from the cavity is sent to a balanced homodyne
detector, measuring the optical quadrature $R_{\theta}^{\text{out}}$
parametrized by the local oscillator (LO) phase $\theta$. The feedback
circuit processes the measured signal according to a specified gain
function $\mu(t)$ and applies a feedback force ($\propto \mu\left(t\right)*R_{\theta}^{\text{out}}\left(t\right)$, where $*$ denotes a convolution)
on the mechanical oscillator accordingly. We restrict ourselves to
feedback that depends linearly on the measurement record, so as to
obtain linear effective equations of motion. In particular this accommodates
the simulation of a viscous force and hence cooling of the mechanical
motion can be engineered.

We seek an effective description of the aforementioned setup including
the feedback mechanism. In the limiting case of Markovian feedback ($\mu(t)\propto\delta(t)$) the dynamics can be described by means of the well developed formalism of feedback
master equations \cite{Wiseman1994,Jacobs2014}. Using these methods, Markovian feedback cooling in the regime of strong quantum cooperativity of optomechanics using variational measurement has been explored in \cite{Hofer2015}. The experimentally more relevant case of Non-Markovian feedback in linear system dynamics is commonly described in the formalism of Heisenberg-Langevin equations which has been fruitfully applied in the context of optomechanics \cite{ManciniVitaliTombesi1998,Genes2008}. We will follow this path in the present article. The basic treatment given in this and the next section will largely reproduce the approach of Genes et al. \cite{Genes2008}.

For our specific purposes of efficient optical readout of the mechanical
motion, linear interaction is in fact well suited. The derivation
of linear optomechanics from the radiation-pressure Hamiltonian is
well-known in the community and here we will largely take the linearized
equations as our starting point (see, e.g., Ref.~\cite{Aspelmeyer2013}
for further details). Essentially, this approach relies on the assumption
that the applied laser drive will induce a large intra-cavity field
amplitude. Henceforth, we consider the dynamics of the optical and
mechanical excursions relative to the corresponding classical steady-state
response. We denote these relative coordinates $X_{\text{m}}$, $P_{\text{m}}$
for the mechanical position and momentum, and $\delta a$ for the
cavity mode amplitude. The linear coupling between these shifted variables
of mechanical motion and light field will be enhanced due to the drive,
which is essential for performing an efficient optical position measurement.
Working in terms of the relative dynamical variables and neglecting
nonlinear terms, as they are not enhanced by the strong driving field,
the linear Heisenberg-Langevin equations for the fluctuations emerge.
Considering resonant readout, where the drive field frequency
$\omega_{\text{d}}$ is aligned with the steady-state cavity resonance,
$\omega_{\text{d}}=\omega_{\text{c}}$, the Heisenberg-Langevin equations
for linear optomechanics including feedback are,
\begin{eqnarray}
\dot{X}_{\text{m}}&=&\omega_{\text{m}}P_{\text{m}},\label{eq:motion eq. position}\\
\dot{P}_{\text{m}}&=&-\omega_{\text{m}}X_{\text{m}}-\gamma_{\text{m}}P_{\text{m}}+g_{\text{om}}X_{\text{c}}+\xi+F_{\text{fb}},\label{eq:motion eq. momentum}\\
\dot{X}_{\text{c}}&=&-\frac{\kappa}{2} X_{\text{c}}+\sqrt{\kappa}X_{\text{c}}^{\text{in}},\label{eq:motion eq. ampl.}\\
\dot{Y}_{\text{c}}&=&-\frac{\kappa}{2} Y_{\text{c}}+g_{\text{om}}X_{\text{m}}+\sqrt{\kappa}Y_{\text{c}}^{\text{in}},\label{eq:motion eq. phase}
\end{eqnarray}
where the optical annihilation operator $\delta a$ is replaced by
cavity amplitude and phase quadratures in an appropriate rotating
frame, $X_{\text{c}}=(e^{i\omega_{\text{c}}t}\delta a+e^{-i\omega_{\text{c}}t}\delta a^{\dagger})/\sqrt{2}$
and $Y_{\text{c}}=(e^{i\omega_{\text{c}}t}\delta a-e^{-i\omega_{\text{c}}t}\delta a^{\dagger})/i\sqrt{2}$.
Eqs.~(\ref{eq:motion eq. position},\ref{eq:motion eq. momentum})
represent the mechanical oscillator of resonance frequency $\omega_{\text{m}}$
and intrinsic damping rate $\gamma_{\text{m}}$, whereas Eqs.~(\ref{eq:motion eq. ampl.},\ref{eq:motion eq. phase})
describe an optical cavity mode which is read out at a rate $\kappa$
(intrinsic cavity damping is equivalent to an imperfect detection
efficiency, which will be introduced later). The coupling between
these two subsystems is characterized by the optomechanical coupling
rate $g_{\text{om}}=\sqrt{\hbar/m\omega_{\text{m}}}(d\omega_{\text{c}}/dx_{\text{m}})\sqrt{2\Phi_{\text{in}}/\kappa}$,
where $\Phi_{\text{in}}$ is the photon flux impinging on
the cavity, $m$ is the effective mass of the mechanical mode and
$(d\omega_{\text{c}}/dx_{\text{m}})$ is the optical frequency shift
per mechanical displacement. Hence, the coupling rate $g_{\text{om}}$
can be tuned via $\Phi_{\text{in}}$ by changing the laser drive power.
Note that the Eqs.~(\ref{eq:motion eq. ampl.},\ref{eq:motion eq. phase})
for the optical quadratures are decoupled from one another due to
the choice of on-resonant driving, $\omega_{\text{d}}=\omega_{\text{c}}$.
In this case the mechanical motion is seen to be read out exclusively
into the phase quadrature, $Y_{\text{c}}$, while the back-action
force on the mechanical mode, $g_{\text{om}}X_{\text{c}}$, comes
entirely from the amplitude fluctuations, $X_{\text{c}}^{\text{in}}$,
as follows from Eq.~(\ref{eq:motion eq. ampl.}).

We now comment on the source terms in Eqs.~(\ref{eq:motion eq. momentum}-\ref{eq:motion eq. phase})
driving the mechanical and optical modes. The feedback force on the
mechanical oscillator is represented by the operator $F_{\text{fb}}$
appearing in Eq.~(\ref{eq:motion eq. momentum}), and we will return to this below. Meanwhile, the thermal noise due to intrinsic mechanical damping,
represented by the Langevin operator $\xi$, can for our purposes
be characterized in the high-temperature limit $k_{\text{B}}T\gg\hbar\omega$
by the following correlation function,
\begin{eqnarray}
\left\langle \xi\left(t\right)\xi\left(t'\right)\right\rangle \approx  \gamma_{\text{m}}(2\bar{n}+1)\delta\left(t-t'\right)\label{eq:thermal correlation}\\
\bar{n} \approx k_{\text{B}}T/\hbar\omega_{\text{m}},\label{eq:phonon number}
\end{eqnarray}
where $\bar{n}$ is the mean number of phonons in the mechanical oscillator
in thermal equilibrium (see Ref.~\cite{GiovannettiVitali2001} for
a discussion of the limitations of this approximation). Turning to
the optical subsystem, we assume the optical amplitude and phase inputs
$X_{\text{c}}^{\text{in}},Y_{\text{c}}^{\text{in}}$ to represent
vacuum fluctuations, i.e., that these operators have the thermal expectation
values
\begin{eqnarray}
\left\langle X_{\text{c}}^{\text{in}}\left(t\right)X_{\text{c}}^{\text{in}}\left(t'\right)\right\rangle & =\left\langle Y_{\text{c}}^{\text{in}}\left(t\right)Y_{\text{c}}^{\text{in}}\left(t'\right)\right\rangle \\ \nonumber \label{eq:vacuum-correl}
& = (1/2)\delta\left(t-t'\right). \\ \nonumber
\end{eqnarray}

To describe the optical readout, we must consider the input-output
relations of the cavity output field,
\begin{eqnarray}
X_{\text{c}}^{\text{out}} & = & \sqrt{\kappa}X_{\text{c}}-X_{\text{c}}^{\text{in}}\nonumber \\
Y_{\text{c}}^{\text{out}} & = & \sqrt{\kappa}Y_{\text{c}}-Y{}_{\text{c}}^{\text{in}}.\label{eq:IO-rel-XY}
\end{eqnarray}
By adjusting the phase of the local oscillator $\theta$ we will be
able to combine the amplitude and phase quadratures in different ratios
with the aim of better balancing measurement error and disturbance.
By considering a general quadrature,
\[
R_{\text{\ensuremath{\theta}}}^{\text{out}(in)}\equiv\cos\theta X_{\text{c}}^{\text{out}(\text{in})}+\sin\theta Y_{\text{c}}^{\text{out}(\text{in})},
\]
the corresponding input-output relation reads
\begin{equation}
R_{\theta}^{\text{out}}=\sqrt{\eta}(\sqrt{\kappa}R_{\theta}-R_{\theta}^{\text{in}})-\sqrt{1-\eta}R_{\theta}^{v},\label{eq:output quadrature}
\end{equation}
where we are accounting for internal cavity loss and measurement imperfection
by a net measurement efficiency $\eta$, assumed to admix a vacuum
field $R^{v}$ that is uncorrelated with $R_{\text{c}}^{\text{in}}$.
Solving Eqs.~(\ref{eq:motion eq. ampl.},\ref{eq:motion eq. phase})
in the Fourier domain by using the convention $F\left(t\right)=\frac{1}{2\pi}\int_{-\infty}^{+\infty}f\left(\omega\right)e^{-i\omega t}d\omega$,
we obtain
\begin{equation}
R_{\theta}\left(\omega\right)=\frac{\sqrt{\kappa}}{\kappa/2-i\omega}R_{\theta}^{{\rm in}}\left(\omega\right)+\frac{\sin\theta\,g_{{\rm om}}}{\kappa/2-i\omega}X_{{\rm m}}\left(\omega\right),\label{eq: Fourier domain of 3,4}
\end{equation}
so that by substituting this into Eq.~(\ref{eq:output quadrature}), the general
output quadrature reads
\begin{eqnarray}
R_{\theta}^{\text{out}}\left(\omega\right)=&\sqrt{\eta}\sin\theta\frac{\sqrt{\kappa}g_{{\rm om}}}{\kappa/2-i\omega}X_{{\rm m}}\left(\omega\right) \nonumber \\
&+\sqrt{\eta}\frac{\kappa/2+i\omega}{\kappa/2-i\omega}R_{\theta}^{\text{in}}\left(\omega\right)-\sqrt{1-\eta}R_{\theta}^{v}\left(\omega\right).\label{eq:new general output}
\end{eqnarray}
making manifest the readout of the position of resonator $X_{{\rm m}}$,
whereas the other terms show the contribution due to measurement noise.
Ignoring the dependence on the parameters of the homodyne measurement
$\theta$ and $\eta$, the maximal rate at which the mechanical motion $X_{{\rm m}}$
can be mapped to the optical quadrature $R_{\theta}^{{\rm out}}$ is given by the ideal measurement rate,
\begin{equation}
\Gamma_{{\rm meas}}=\frac{4g_{{\rm om}}^{2}}{\kappa},\label{eq:measurement rate}
\end{equation}
which in the bad-cavity limit is the square of the
coefficient mapping the mechanical oscillator position into the optical readout in~(\ref{eq:new general output}) for $\eta=1$ and $\theta=\pi/2$. For other values of $\eta,\theta$, the effective readout rate is reduced by a factor of $\eta \sin^2\theta$.

Finally, we address the relationship between the feedback force $F_\mathrm{fb}$ and the optical homodyne measurement
of $R_{\text{\ensuremath{\theta}}}^{\text{out}}$. The feedback circuit
integrates the measured quadrature signal up to the present time $t$.
Since we are interested in preserving the linearity of the equations
of motion, (\ref{eq:motion eq. position}-\ref{eq:motion eq. phase}),
we take the feedback force to be given by a temporal convolution
\begin{equation}
F_{\text{fb}}(t)=-\int_{-\infty}^{t}ds\mu(t-s)R_{\theta}^{\text{est}}(s),\label{eq:Feedback operature}
\end{equation}
where $\mu\left(\tau\right)=\mathcal{F}^{-1}\{\mu(\omega)\}$ is the inverse
Fourier transform of the spectral gain function, $\mu(\omega)$, and
\begin{equation}
R_{\theta}^{\text{est}}\equiv\frac{R_{\text{\ensuremath{\theta}}}^{\text{out}}}{\sqrt{\kappa\eta}\sin\theta}=\frac{1}{\sqrt{\kappa\eta}}(\cot\theta X_{\text{c}}^{\text{out}}+Y_{\text{c}}^{\text{out}})\label{eq:estimated measrement quadrature}
\end{equation}
is the rescaled measurement quadrature. The local oscillator phase
$\theta$ in the homodyne measurement selects which light quadrature
to condition the feedback force on, as was mentioned above. That we
convolve with $R_{\theta}^{\text{est}}$ in Eq.~(\ref{eq:Feedback operature})
rather than with the original $R_{\theta}^{\text{out}}$
of Eq.~(\ref{eq:output quadrature}) amounts to a scaling convention
for the gain function $\mu(t)$. The convention used here is designed
to remove the terms which are related to the measurement apparatus ($\sqrt{\eta}\sin\theta$)
from the proportionality factor in the relationship $R_{\text{\ensuremath{\theta}}}^{\text{out}}\propto X_{\text{m}}$ in Eq.~(\ref{eq:new general output}).
To be clear, $\mu(t)$ is the gain applied after having corrected for the
measurement inefficiency $\eta$ and the quadrature angle entering
as $\sin\theta$. This choice allows us to vary $\theta$ while keeping
fixed the net gain of the position component of the measurement, in
turn, keeping the feedback-induced damping fixed as we will see below.

\section{Mechanical response and effective susceptibility \label{sec:Mech-resp}}

Having established the equations of motion (\ref{eq:motion eq. position}-\ref{eq:motion eq. phase})
and the relevant optical input-output relation (\ref{eq:new general output}),
we turn to solving this set of equations for the mechanical response.
Since the system is linear, this is straightforwardly done in the
Fourier domain. A useful way of expressing the solution for the mechanical
mode is the response relation (suppressing the $\omega$ dependence
of the source terms for brevity)

\begin{equation}
X_{\text{m}}\left(\omega\right)=\chi_{\text{eff}}(\omega)[\xi+f_{\text{ba}}+f_{\text{fb}}+f_{v}],\label{eq:response relation}
\end{equation}
where the effective mechanical susceptibility is

\begin{equation}
\chi_{\text{eff}}^{-1}(\omega)=\frac{1}{\omega_{\text{m}}}[\omega_{\text{m}}^{2}-\omega^{2}-i\gamma_{\text{m}}\omega+\mu(\omega)\frac{g_{\text{om}}\omega_{\text{m}}}{\kappa/2-i\omega}],\label{eq:effective susseptibility}
\end{equation}
and the four stochastic forces driving the oscillator are the thermal
Langevin operator $\xi$ and the fluctuation associated with back-action,
feedback, and extraneous vacuum:

\begin{equation}
f_{\text{ba}}(\omega)=\frac{\sqrt{\kappa}g_{\text{om}}}{\kappa/2-i\omega}X_{\text{c}}^{\text{in}}(\omega),\label{eq:back action noise}
\end{equation}
\begin{equation}
f_{\text{fb}}(\omega)=-\frac{\kappa/2+i\omega}{\kappa/2-i\omega}\frac{\mu(\omega)}{\sqrt{\kappa}}\left[\cot\theta X_{\text{c}}^{\text{in}}(\omega)+Y_{\text{c}}^{\text{in}}(\omega)\right],\label{eq:feedback noise}
\end{equation}

\begin{equation}
f_{v}(\omega)=\frac{\mu(\omega)}{\sqrt{\kappa}}\sqrt{\eta^{-1}-1}\left[\cot\theta X^{v}(\omega)+Y^{v}(\omega)\right].\label{eq:vacuum noise}
\end{equation}
The back-action force $f_{\text{ba}}$ arises from the optical amplitude
fluctuations $X_{\text{c}}^{\text{in}}$, whereas the noise of the
meter field $f_{\text{fb}}$ introduced via the feedback force contains
both optical amplitude and phase fluctuations, $X_{\text{c}}^{\text{in}}$
and $Y_{\text{c}}^{\text{in}}$, whereby $f_{\text{ba}}$ and $f_{\text{fb}}$
are correlated (for general $\theta$). $f_{v}$ is the part of the
feedback force fluctuations coming from extraneous vacuum noise due
to optical losses.

We see that the feedback gain function $\mu(\omega)$, i.e. the Fourier
transform of $\mu(\tau)$, enters in the effective mechanical susceptibility
$\chi_{\text{eff}}$ as well as $f_{\text{fb}}$ and $f_{v}$, Eqs.
(\ref{eq:effective susseptibility},\ref{eq:feedback noise},\ref{eq:vacuum noise}).
The choice of the function $\mu(\omega)$ can hence be thought of as
influencing both the response characteristics and the spectral mapping
of imprecision noise into the oscillator. Following Ref.~\cite{Genes2008}
we choose the gain function

\begin{equation}
\mu(t)=\mu_{\text{fb}}\frac{d}{dt}[\Theta(t)\omega_{\text{fb}}e^{-t\omega_{\text{fb}}}],\label{eq:gain time}
\end{equation}
which in the frequency domain becomes
\begin{equation}
\mu(\omega)=\frac{-i\omega \mu_{\text{fb}}}{1-i\frac{\omega}{\omega_{\text{fb}}}},\label{eq:gain frequency}
\end{equation}
where $\mu_{\text{fb}}$ is the dimensionless feedback gain, $\omega_{\text{fb}}$
is a low-pass cut-off frequency, and $\Theta(t)$ is the Heaviside
function. Considering Eq.~(\ref{eq:gain frequency}) in the limit
$\omega_{\text{fb}}\rightarrow\infty$, we see that it corresponds
to a time derivative of the measurement current. The rationale behind this choice is to achieve an effective friction force  ($F_{fb}\propto-P_{m}\propto-\dot X_m$) which is attained through the time derivative of the photocurrent given the fact that light reads out the position of the mechanical oscillator. For finite $\omega_{\text{fb}}$,
we can then think of the feedback filter as performing a derivative
combined with a low-pass filter. Note, however, that the value of
$\omega_{\text{fb}}$ not only controls the frequency-cutoff of $|\mu(\omega)|$,
but also influences the feedback phase function $\text{Arg}[\mu(\omega)]$.
By considering Eq.~(\ref{eq:effective susseptibility}) for this choice
of feedback function, Eq.~(\ref{eq:gain frequency}), we can extract the
effective mechanical resonance frequency and dissipation rate. In
the bad-cavity limit we find
\begin{eqnarray}
\omega_{\text{eff,m}}(\omega)&=&\sqrt{\omega_{\text{m}}^{2}+\frac{\sigma\gamma_{\text{m}}\omega_{\text{fb}}\omega^{2}}{\omega_{\text{fb}}^{2}+\omega^{2}}},\label{eq:effective frequency large-kappa}\\
\gamma_{\text{eff,m}}(\omega)&=&\gamma_{\text{m}}\left(1+\frac{\sigma\omega_{\text{fb}}^{2}}{\omega_{\text{fb}}^{2}+\omega^{2}}\right),\label{eq:effective dissipation large-kappa}\\
\chi_{\text{eff}}^{-1}(\omega)&=&\frac{1}{\omega_{\text{m}}}[\omega_{\text{eff,m}}^{2}(\omega)-\omega^{2}-i\gamma_{\text{eff,m}}(\omega)\omega],\label{eq:Susseptibility large-kappa}
\end{eqnarray}
where
\begin{equation}
\sigma\equiv\frac{2\mu_{\text{fb}}g_{\text{om}}\omega_{\text{m}}}{\kappa\gamma_{\text{m}}}\label{eq:sigma-def}
\end{equation}
is the rescaled dimensionless feedback gain. More precisely, then, we define the bad-cavity limit as $\kappa\rightarrow\infty$ while keeping $\sigma$ and $\Gamma_{{\rm meas}}$ finite. Practical considerations
may limit the available range of the \emph{absolute} feedback gain
$\tilde{\sigma}=\sigma/(\sqrt{\kappa\eta}\sin\theta)$, cf. Eqs.~(\ref{eq:Feedback operature},\ref{eq:estimated measrement quadrature}). Deviation from $\eta=1$ and $\theta=\pi/2$ requires larger absolute gain $\tilde{\sigma}$ in order to maintain
a certain value of $\sigma$, the gain parameter entering $\chi_{\text{eff}}(\omega)$.
The stability of the optomechanical system in the presence of feedback
can be determined from the complex poles of $\chi_{\text{eff}}(\omega)$
using the Routh-Hurwitz criterion \cite{DeJesus1987}. Here it suffices
to remark that in the idealized bad-cavity limit considered here,
$\kappa\rightarrow\infty$, the stability criterion is fulfilled for
all values of $\sigma$. (We give the stability criterion for arbitrary
values of $\kappa$ in \ref{sec:stability}.)

\section{Variational measurements and Heisenberg's microscope\label{sec:Heisenberg-microscope}}

Before calculating the mechanical steady-state occupation from the
solution found in the preceding section, it is appropriate to pause
for a more qualitative discussion of the physics that will emerge
from the analysis. At a very basic level, the idea of feedback cooling
of a mechanical oscillator is simply that if we monitor its motion
(by means of some meter degree of freedom), we may, based on this
information, apply an effectively viscous force, that will dampen
the motion. The extent to which we are able to bring the motion to
a halt by this technique will be determined by the interplay of (at
least) four effects as seen from Eq.~(\ref{eq:response relation}):
Firstly, the thermal noise, $\xi$, driving the motion due to the
internal friction mechanisms of the mechanical element. Secondly,
the disturbing back-action force, $f_{\text{ba}}$, of the meter system
on the mechanical motion, e.g., the radiation-pressure shot noise
arising from the random timing of the momentum kicks imparted on the
mechanical oscillator by the impinging photons. Thirdly, the imprecision
noise, $f_{\text{fb}}$, of the position measurement, which limits
the ability to apply the right amount of force required to halt the
motion. Fourthly, the feedback modification of the mechanical response
function, $\chi_{\text{eff}},$ by inducing an increased damping.

In considering how to balance these effects, it is important to acknowledge
the time-continuous nature of the scheme: For instance, if we were
to attempt a perfectly precise instantaneous position measurement,
the back-action force would introduce a large uncertainty in the mechanical
momentum, as per Heisenberg's uncertainty relation, rendering the
position a short while later completely unpredictable. If, on the
other hand, we were to measure very weakly to avoid disturbing the
system, the imprecision noise would dominate and very little information
would be gained. Both outcomes are clearly at odds with the desired
goal of cooling and we are therefore led to strike a balance between
the influences of back-action and imprecision noise. In terms of the
Gedankenexperiment of Heisenberg's microscope, this trade-off would
correspond to sacrificing (instantaneous) position resolution to gain
increased information about the direction of the scattered photon
(see Fig.~\ref{fig:Heisenberg microscope-simple}). The possibility
of further optimizing feedback cooling in this way has previously
been pointed out in Ref.~\cite{Genes200933}, although without extensive
analysis or discussion. In the remainder of this section, we will
provide intuition as to why variational measurements can be advantageous.

While we must generally include the thermal influence of intrinsic
mechanical damping in the analysis (and will do so below), the main
interest of this work is the regime of quantum operation, where this
thermal load is perturbative. To establish intuition that will be
useful in interpreting the results of the rigorous analysis to be
presented in subsequent sections, we therefore proceed now to consider
the trade-off between back-action and imprecision noise. This simpler
scenario, in which we only consider the fundamental fluctuations required
by quantum mechanics, is only adequate to the extent that
the measurement and feedback processes occur fast compared to the thermal coherence time $\sim 1/\gamma_{\text{m}}\bar{n}$.
This is the limit of very large quantum cooperativity,
\[
C_{\text{q}}\equiv\frac{\Gamma_{\text{meas}}}{\gamma_{\text{m}}\bar{n}}=\frac{4g_{\text{om}}^2}{\kappa\gamma_{\text{m}}\bar{n}}\gg1,
\]
and fast feedback $\sigma\propto\Gamma_{\text{meas}}/\gamma_{\rm m}$ (whereas
in current state-of-the-art experiments, $C_{\text{q}}\gtrsim1$).
In this limit we may for the purposes of the present discussion take
the mechanical response, (\ref{eq:response relation}), to be (in
the bad-cavity limit and assuming perfect detection, $\eta=1$)
\begin{eqnarray}
X_{\text{m}}(\omega)  &=  \chi_{\text{eff}}(\omega)F(\omega)\label{eq:mech-X-response_F}\\
F(\omega) & \approx f_{\text{ba}}(\omega)+f_{\text{fb}}(\omega) \nonumber\\ &=\left[\sqrt{\Gamma}_{\rm meas}-\frac{\mu(\omega)}{\sqrt{\kappa}}\cot\theta\right]X_{\text{c}}^{\text{in}}(\omega) \nonumber \\
& -\frac{\mu(\omega)}{\sqrt{\kappa}}Y_{\text{c}}^{\text{in}}(\omega),\label{eq:F-highCq}
\end{eqnarray}
where the expression for the force $F(\omega)$ is organized according
to its contributions from amplitude and phase fluctuations. Eq.~(\ref{eq:F-highCq})
emphasizes the fact that for general $\theta$ the amplitude fluctuations
drive the mechanical motion both directly, via the back-action force
$f_{\text{ba}}$, and indirectly, via the fluctuations $f_{\text{fb}}$
injected by the feedback mechanism. As these contributions add coherently
in determining the mechanical response, the possibility of destructive
interference arises. Moreover, Eq.~(\ref{eq:F-highCq}) shows that
the interference varies with $\omega$. This dependence must be considered
over the effective mechanical bandwidth, which is typically set by
the feedback-induced broadening, Eq.~(\ref{eq:effective dissipation large-kappa}).
This observation hints at a trade-off between, on the one hand, achieving
favorable interference over the entire effective bandwidth and, on
the other hand, suppressing thermal noise by applying a large feedback
gain.

Having established the interference effect between back-action and
feedback forces, we now turn to the question of what constitutes favorable
interference and, in particular, how $\theta$ and $\mu(\omega)$ should
be chosen to attain this. In the classical regime, where thermal noise
dominates, back-action noise can be neglected and the optimal measurement
quadrature is the phase quadrature ($\theta=\pi/2$), into which the
position measurement is read out (as described in Sec.~\ref{sec:Model}).
In the quantum regime this is no longer the case as can be demonstrated
by a simple geometrical argument, that we will now turn to. Since
the purpose of the scheme is to map the vacuum state of light onto
the mechanical mode, as mentioned previously, the scheme can only
be successful if the (orthogonal) noise quadratures of light $X_{\text{c}}^{\text{in}},Y_{\text{c}}^{\text{in}}$
are mapped to orthogonal mechanical quadratures with equal strength.
If this were not the case, it would violate equipartition between
the mechanical quadratures and thus the resulting mechanical state
could impossibly be the ground state. To understand the mapping into
$X_{\text{m}}$ and $P_{\text{m}}$, we note that the Fourier transform
of Eq.~(\ref{eq:motion eq. position}) is $P_{\text{m}}(\omega)=-i(\omega/\omega_{\text{m}})X_{\text{m}}(\omega)$.
The relative phase of $(-i)$ between the position and momentum response
entails that the real and imaginary parts of $X_{\text{m}}(\omega)=\chi_{\text{eff}}(\omega)F(\omega)$, which are
$X_{\text{m}}(\omega)+X_{\text{m}}^{\dagger}(\omega)$ and $[X_{\text{m}}(\omega)-X_{\text{m}}^{\dagger}(\omega)]/i$,
will map to orthogonal mechanical quadratures as seen from the time-domain
response (in a narrow-band approximation for simplicity)
\begin{eqnarray}
X_{\text{m}}(t) & \propto & \left(e^{-i\omega_{\text{m}}t}X_{\text{m}}(\omega_{\text{m}})+\text{H.c.}\right)\nonumber\\
 & = &  \cos(\omega_{\text{m}}t)\left[X_{\text{m}}(\omega_{\text{m}})+X_{\text{m}}^{\dagger}(\omega_{\text{m}})\right] \nonumber\\
& &+\sin(\omega_{\text{m}}t)\left[X_{\text{m}}(\omega_{\text{m}})-X_{\text{m}}^{\dagger}(\omega_{\text{m}})\right]/i
\end{eqnarray}
\begin{eqnarray}
P_{\text{m}}(t) & \propto & \left(e^{-i\omega_{\text{m}}t}P_{\text{m}}(\omega_{\text{m}})+\text{H.c.}\right)\nonumber\\
 & \approx & \cos(\omega_{\text{m}}t)\left[X_{\text{m}}(\omega_{\text{m}})-X_{\text{m}}^{\dagger}(\omega_{\text{m}})\right]/i\nonumber\\
 & & -\sin(\omega_{\text{m}}t)\left[X_{\text{m}}(\omega_{\text{m}})+X_{\text{m}}^{\dagger}(\omega_{\text{m}})\right].
\end{eqnarray}
Thus, the above equipartition argument requires $X_{\text{m}}(\omega_{\text{m}})+X_{\text{m}}^{\dagger}(\omega_{\text{m}})$
and $[X_{\text{m}}(\omega_{\text{m}})-X_{\text{m}}^{\dagger}(\omega_{\text{m}})]/i$
to represent orthogonal quadratures of light with equal weight.

\begin{figure}[H]
\begin{centering}
\includegraphics[width=1\columnwidth]{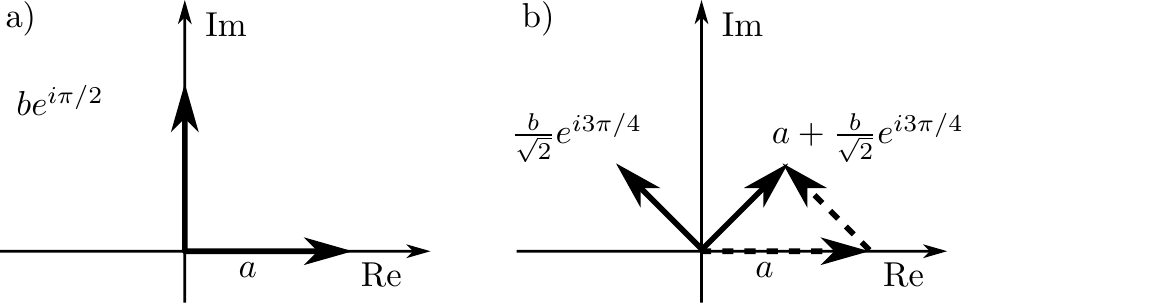}

\caption{Geometrical considerations regarding the
choice of homodyne quadrature angle $\theta$. Depicted are the mapping coefficients of $X_{\text{c}}^{\text{in}}(\omega),Y_{\text{c}}^{\text{in}}(\omega)$ in~(\ref{eq:F-net_geom-1}) as solid arrows in the complex plane. Two cases are considered:
a) $\phi=\pi/2,\theta=\pi/2,b=a$; b) $\phi=3\pi/4,\theta=\pi/4,b=a/\sqrt{2}$.\label{fig:Geometrical-considerations-1}}
\end{centering}
\end{figure}

Let us now apply this mapping condition to Eqs.~(\ref{eq:mech-X-response_F},\ref{eq:F-highCq})
near resonance $\omega\approx\omega_{\text{m}}$. We introduce the
simplified symbols $a\equiv\sqrt{\Gamma}_{\rm meas}$, $b\equiv|\mu(\omega_{\text{m}})/\sqrt{\kappa}|,$
and $\phi=\text{Arg}[-\mu(\omega_{\text{m}})/\sqrt{\kappa}]$, whereby
the force, (\ref{eq:F-highCq}), can be expressed as
\begin{equation}
F(\omega)=(a+be^{i\phi}\cot\theta)X_{\text{c}}^{\text{in}}\left(\omega\right)+be^{i\phi}Y_{\text{c}}^{\text{in}}\left(\omega\right).\label{eq:F-net_geom-1}
\end{equation}
We now fix $a$ and consider two characteristic values of $\phi$
(see Fig.~\ref{fig:Geometrical-considerations-1}): For $\phi=\pi/2$,
$f_{\text{ba}}$ and $f_{\text{fb}}$ add in quadrature and we see
from (\ref{eq:F-net_geom-1}) that matching can only be achieved for
$\theta=\pi/2$ and $b=a$ as illustrated in Fig.~\ref{fig:Geometrical-considerations-1}a.
Consider now the case of $\phi=3\pi/4$, whereby part of $f_{\text{fb}}$
is anti-correlated with $f_{\text{ba}}$ and matching can be achieved
by choosing $\theta=\pi/4$ and $b=a/\sqrt{2}$ (Fig.~\ref{fig:Geometrical-considerations-1}b).
In the second scenario, part of the direct back action is cancelled
by partially measuring the amplitude fluctuations in the light field
and feeding them back into the mechanical mode via the feedback force,
while operating at the same absolute feedback gain $\propto b/\sin\theta$
as in the first scenario (with $\theta=\pi/2$). Hence the second
scenario leads to less fluctuations in the mechanical mode and thus
the resulting state will be closer to the ground state as will in
fact be found in the rigorous analysis below. These simple considerations
indicate that in the quantum regime it is important to properly choose
both the homodyne measurement quadrature, via $\theta$, and the feedback
gain $\mu(\omega)$. Note however, as remarked below Eq.~(\ref{eq:F-highCq}),
that the matching consideration illustrated in Fig.~\ref{fig:Geometrical-considerations-1}
should be applied to all frequencies within the effective mechanical
bandwidth. Since our chosen value of $\theta$ is a constant, whereas
the feedback gain $\mu(\omega)$ is frequency dependent, it will in
general not be possible to achieve advantageous interference over
the entire bandwidth. In the next section we resume the quantitative
mathematical analysis.

\section{Steady-state occupation in the bad-cavity regime\label{sec:Steady-state-occupation}}

Given the solution (\ref{eq:response relation}) for the spectral
response of the mechanical mode, we are now in a position to calculate
the average number of phonons $n_{\text{tot}}$ it contains in the
steady state of the feedback scheme. This quantity will serve as our
figure of merit, with $n_{\text{tot}}<1$ being the regime of interest.
By considering the Hamiltonian of the free mechanical oscillator,

\begin{equation}
H=\frac{\hbar\omega_{\text{m}}}{2}\left( X_{\text{m}}^{2}+ P_{\text{m}}^{2} \right)=\hbar\omega_{\text{m}}\left(\hat{n}_{\text{tot}}+\frac{1}{2}\right),\label{eq:Hamiltonian of oscillator}
\end{equation}
which is stated in terms of the variance of the dimensionless operators
$X_{\text{m}}$ and $P_{\text{m}}$, we can extract the mean number
of phonons ${n}_{\text{tot}}=\langle\hat{n}_{\text{tot}}\rangle$ as an integral over spectral density of each quadrature,
\begin{equation}
n_{\text{tot}}=\frac{1}{2}\int_{-\infty}^{\infty}\frac{d\omega}{2\pi}\left(S_{X}\left(\omega\right)+S_{P}\left(\omega\right)\right)-\frac{1}{2},\label{eq:Spectral integral}
\end{equation}
where we have introduced the position and momentum fluctuation spectra
\begin{eqnarray}
S_{X}\left(\omega\right) & = & \int_{-\infty}^{\infty}d\omega'\left\langle X_{\text{m}}\left(\omega\right)X_{\text{m}}\left(\omega'\right)\right\rangle ,\label{eq:Position spectrum}\\
S_{P}\left(\omega\right) & = & \int_{-\infty}^{\infty}d\omega'\left\langle P_{\text{m}}\left(\omega\right)P_{\text{m}}\left(\omega'\right)\right\rangle .\label{eq:Momentum spectrum}
\end{eqnarray}
Using Eq.~(\ref{eq:response relation}) the position spectrum reads
\begin{eqnarray}
S_{X}\left(\omega\right) = & \int_{-\infty}^{\infty}d\omega'|\chi_{\text{eff}}(\omega)|^{2}\bigg(\left\langle \xi(\omega)\xi(\omega')\right\rangle \nonumber\\ & +\left\langle f_{\text{ba}}(\omega)f_{\text{ba}}(\omega')\right\rangle +\left\langle f_{\text{fb}}(\omega)f_{\text{fb}}(\omega')\right\rangle \nonumber\\
  &+\left\langle f_{\text{co}}(\omega)f_{\text{co}}(\omega')\right\rangle +\left\langle f_{v}(\omega)f_{v}(\omega')\right\rangle \bigg),\label{eq:Position spectrum in separated form}
\end{eqnarray}
where
\begin{equation}
\left\langle f_{\text{co}}(\omega)f_{\text{co}}(\omega')\right\rangle \equiv\left\langle f_{\text{fb}}\left(\omega\right)f_{\text{ba}}\left(\omega'\right)+f_{\text{ba}}\left(\omega\right)f_{\text{fb}}\left(\omega'\right)\right\rangle
\end{equation}
is the correlation between back-action noise and feedback force. As
discussed in Sec.~\ref{sec:Heisenberg-microscope}, the indirect back
action, coming through the feedback force, is responsible for this
cross correlation. As one can easily find by using the Fourier transform
of Eq.~(\ref{eq:motion eq. position}), we use the relation $S_{P}=\omega^{2}S_{X}/\omega_{\rm m}^{2}$
to simplify the calculation of Eq.~(\ref{eq:Spectral integral}). The
correlation functions appearing on the right-hand side of Eq.~(\ref{eq:Position spectrum in separated form})
can be determined from the definitions in Eqs.~(\ref{eq:back action noise}-\ref{eq:vacuum noise})
by using appropriate Fourier domain equivalents of Eqs.~(\ref{eq:thermal correlation}-\ref{eq:vacuum-correl}).
 By performing the integral (\ref{eq:Spectral integral}) (using
the analytical procedure given in \ref{sec:Rational-Integral})
we obtain an expression of the form

\begin{equation}
n_{\text{tot}}=n_{\text{th}}+n_{\text{ba}}+n_{\text{fb}}+n_{\text{co}}+n_{v}-\frac{1}{2},\label{eq:total number}
\end{equation}
where the various contributions correspond to the respective terms
in Eq.~(\ref{eq:Position spectrum in separated form}). The clear
physical origin of these terms aids the interpretation of their impact
on the total occupation number. The cross-correlation between direct back-action and the feedback noise is represented by $n_{\text{co}}$,
which for parameters of interest is negative.

Since feedback cooling is typically applied in the bad-cavity regime,
$\kappa\gg\omega_{\text{m}},\omega_{\text{fb}}$, we will henceforth
focus on this parameter regime for simplicity of analysis (see the
\ref{sec:Exact-solution-for} for a general expression for
$n_{\text{tot}}$ valid for all values of $\kappa$). We parametrize
the feedback cut-off frequency $\omega_{\text{fb}}$ by its ratio
to the mechanical resonance frequency
\begin{equation}
\alpha\equiv\omega_{\text{fb}}/\omega_{\text{m}},\label{eq:alpha}
\end{equation}
and note that $\alpha$ is typically within an order of magnitude
of unity. Introducing the mechanical quality factor $Q_{\text{m}}\equiv\omega_{\text{m}}/\gamma_{\text{m}}$,
for which values on the order of $10^{6}$ and beyond are routinely achieved
in optomechanical experiments, we can therefore safely make the additional
assumption that $Q_{\text{m}}\gg\alpha,\alpha^{-1}$. Under these
assumptions the various contributions to $n_{\text{tot}}$ (\ref{eq:total number})
are evaluated to
\begin{eqnarray}
n_{\text{th}}&=&\frac{1}{D}\left(\overline{n}+\frac{1}{2}\right)\left(1+\alpha^{-2}+\frac{1}{\alpha}\frac{\sigma}{2Q_{\text{m}}}\right),\label{eq:thermal number}\\
n_{\text{ba}}&=&\frac{C_{\text{cl}}}{4D}\left(1+\alpha^{-2}+\frac{1}{\alpha}\frac{\sigma}{2Q_{\text{m}}}\right),\label{eq:back action number}\\
n_{\text{fb}}&=&\frac{\sigma^{2}}{4 C_{\text{cl}} D}\left(1+\alpha\frac{\sigma}{2Q_{\text{m}}}\right)\csc^{2}\left(\theta\right),\label{eq:feedback number}\\
n_{\text{co}}&=&-\frac{\sigma}{2\alpha D}\left(1+\alpha\frac{\sigma}{2Q_{\text{m}}}\right)\cot\left(\theta\right),\label{eq:correlation number}\\
n_{v}&=&n_{\text{fb}}\left(\eta^{-1}-1\right),\label{eq:vacuum number}\\
D&\equiv&1+\sigma+\alpha^{-2},\label{eq:Simplified denominator}
\end{eqnarray}
where
\begin{equation}
C_{\text{cl}}\equiv4g_{\text{om}}^{2}/\kappa\gamma_{\text{m}}\label{eq:classical cooperativity}
\end{equation}
is the classical cooperativity.

These results are relatively simple and we will now discuss their
behavior. We start by noting that the feedback-induced mechanical
broadening (\ref{eq:effective dissipation large-kappa}) is given
by $\gamma_{\text{m,eff}}(\omega_{\text{m}})-\gamma_{\text{m}}=\sigma/[1+\alpha^{-2}]$,
within a narrow-band approximation $\omega\approx\omega_{m}$. This
increased broadening will tend to decouple the mechanical mode from
its thermal bath while adding low-temperature noise from the optical bath resulting in net cooling, as is seen by considering the
behavior of $n_{\text{th}}$ (\ref{eq:thermal number}) with increasing
$\sigma$. However, this only holds insofar as the scaled cut-off feedback
frequency $\alpha$ is large enough for the feedback mechanism to
be able to react on the appropriate timescale. Moreover, there is
a limit to how much $n_{\text{th}}$ can be suppressed in this way
which manifests itself when the effective mechanical quality factor
becomes too small, $Q_{\text{m,eff}}\equiv\omega_{\text{m}}/\gamma_{\text{m,eff}}\ll\alpha,\alpha^{-1}$,
as can be seen by taking the limit $\sigma\rightarrow\infty$. Since
we treat both the thermal noise and the back-action as white noise
(given the lack of cavity filtering in the bad cavity regime), Eqs.
(\ref{eq:thermal number},\ref{eq:back action number}) are seen to
be very similar, with $C_{\text{cl}}/4$ simply playing the role of $\bar{n}+1/2,$
i.e. an equivalent back-action noise flux per unit bandwidth.

We now turn to the imprecision noise contribution $n_{\text{fb}}$,
(\ref{eq:feedback number}). The increase in
$n_{\text{fb}}$ seen when $\theta$ moves away from $\pi/2$ occurs
because this degrades the signal-to-noise ratio of the homodyne measurement.
Unsurprisingly, $n_{\text{fb}}$ increases with scaled effective gain
$\sigma$, in fact it diverges as expected from amplifying a noisy
measurement excessively. Interestingly, $n_{\text{fb}}$ also diverges
for $\alpha\rightarrow\infty$, i.e., when taking the feedback frequency
cut-off $\omega_{\text{fb}}$ to infinity which yields a derivative
filter, see Eq.~(\ref{eq:gain frequency}). While this is ideal from
the point of view of estimating the instantaneous mechanical velocity,
it simultaneously feeds amplified imprecision noise into the oscillator
from an unbounded spectral range resulting in $n_{\text{fb}}\rightarrow\infty$
(as $\alpha\rightarrow\infty$). This observation prompts us to choose
values of $\alpha$ which are not too large, which runs counter to
the demand of having a large $\alpha$ to be able to cool out the
mechanical noise. Hence the finite optimal value for $\alpha$ somehow
balances these considerations. Imperfect homodyne detection or optical
losses inject additional (uncorrelated) vacuum noise into the measurement
current as accounted for by $n_{v}$, Eq.~(\ref{eq:vacuum number}).

The novel aspect of the present work hinges on the fact that for a
general homodyne quadrature $\theta$, the imprecision shot noise
of the measurement is correlated with the back-action noise on the
mechanical mode as discussed previously. If we have anti-correlations,
$n_{\text{co}}<0$, destructive interference lessens the total mechanical
occupancy $n_{\text{tot}}$ potentially leading to an advantage over
the conventional choice of $\theta=\pi/2$ that maximizes the signal-to-noise
ratio of the measurement. Since the dependence of $n_{\text{tot}}$
on $\theta$ is just that of Eqs.~(\ref{eq:feedback number}-\ref{eq:vacuum number}),
the value of $\theta_{\text{opt}}$ which minimizes $n_{\text{tot}}$
is straightforwardly found to be
\begin{equation}
\theta_{\text{opt}}=\text{Arccot}\left(\frac{\it{C}_{\text{cl}}\eta}{\alpha\sigma}\right).\label{eq:theta-opt}
\end{equation}
Substituting $\theta=\theta_{\text{opt}}$ back into Eqs.~(\ref{eq:feedback number},\ref{eq:correlation number})
and summing all contributions according to~(\ref{eq:total number}),
we find,

\begin{eqnarray}
\left.n_{\text{tot}}\right|_{\theta_{\text{opt}}}= -\frac{1}{2}+\left(1+\sigma+\alpha^{-2}\right)^{-1}\times \nonumber \\
\Bigg[\left(\overline{n}+\frac{1}{2}+\frac{C_{\text{cl}}}{4}\right)\left(1+\alpha^{-2}+\frac{1}{\alpha}\frac{\sigma}{2 Q_{\text{m}}}\right) \nonumber\\
+\left(\frac{\sigma^{2}}{4 C_{\text{cl}}\eta}-\frac{C_{\text{cl}}\eta}{4\alpha^{2}}\right)\left(1+\alpha\frac{\sigma}{2Q_{\text{m}}}\right)\Bigg].\label{eq:solution for optimized theta}
\end{eqnarray}
The optimal values $\theta_{\text{opt}}$ and $n_{\text{tot}}$, (\ref{eq:theta-opt})
and (\ref{eq:solution for optimized theta}), are given as functions
of $C_{\text{cl}}$, $\sigma$, and $\alpha$. These remaining parameters
will be optimized in the next section.

\section{Optimized cooling\label{sec:Optimized-cooling}}

\begin{figure*}[b]
\begin{centering}
\includegraphics[width=1.7\columnwidth]{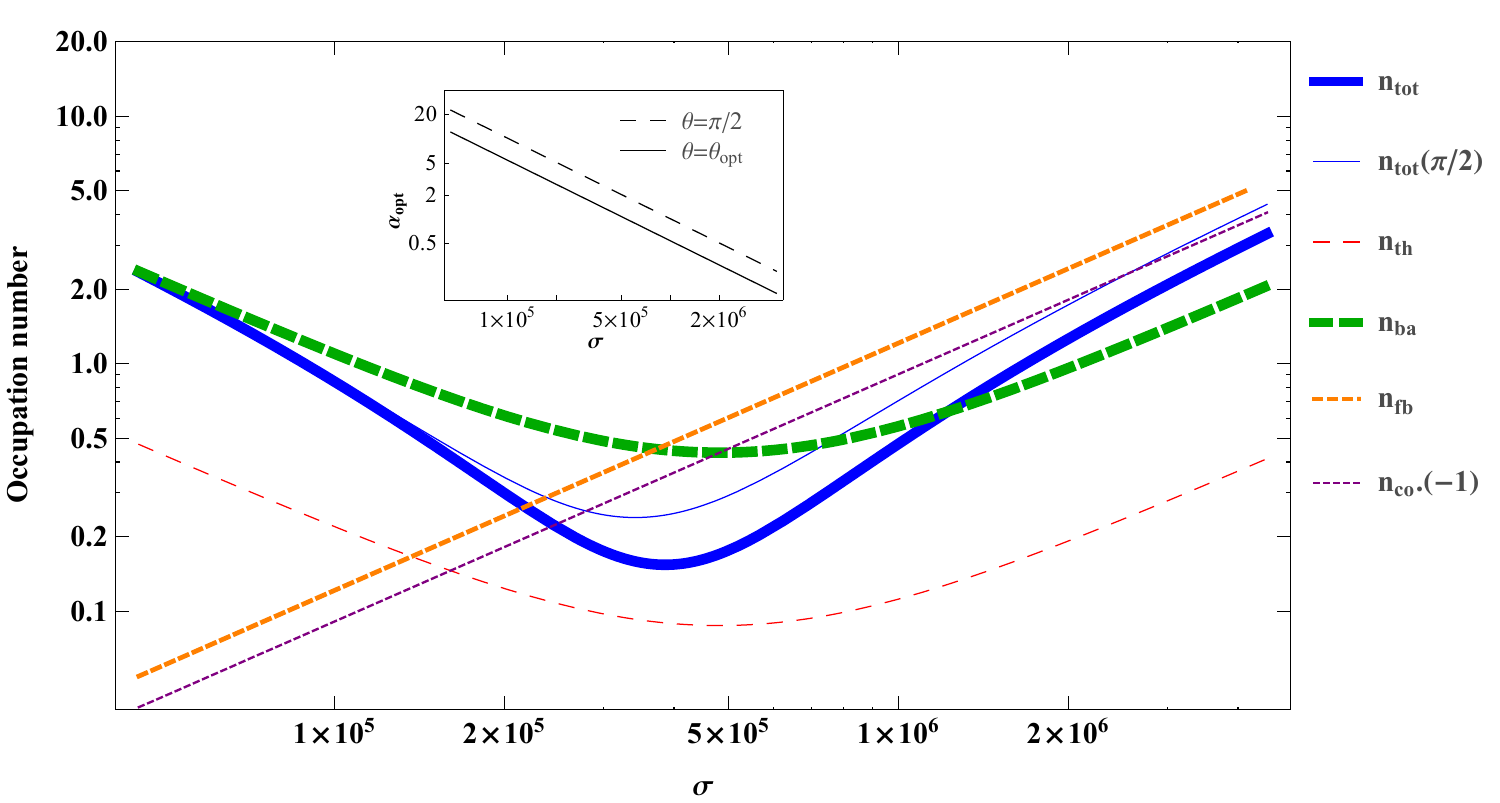}

\caption{Mechanical steady-state occupancy $n_{\text{tot}}$ and its individual
contributions as functions of the feedback strength $\sigma$,
(\ref{eq:sigma-def}), evaluated for the optimized values $\theta_{\text{opt}}(\sigma),\alpha_{\text{opt}}(\sigma)$.
The parameter set used here is $Q_{\text{m}}=10^{6},\bar{n}=2.1\times10^{4}$, which is close to the parameters of \cite{WilsonSudhirPiroEtAl2015}, and $C_{\text{q}}=20$. We neglect losses, $\eta=1$. In addition, $n_{\text{tot}}(\theta=\pi/2)$ is plotted
for comparison, where we have evaluated it at the optimal value of
$\alpha$ given the suboptimal choice $\theta=\pi/2$, $\alpha_{\text{opt}}^{(\pi/2)}(\sigma)$.
The optimal angle is, for the reasons explained in the main text, approximately constant over the plotted range of $\sigma$, $\theta_{\text{opt}}(\sigma)\approx 52{}^{\circ}$, and is set by the value of the cooperativity. (Inset) Plot of the functions
$\alpha_{\text{opt}}(\sigma)$ and $\alpha_{\text{opt}}^{(\pi/2)}(\sigma)$
that minimize $n_{\text{tot}}$ and $n_{\text{tot}}(\theta=\pi/2)$,
respectively.\label{fig:Func-gamma-fb}}
\end{centering}
\end{figure*}

Having derived relatively simple expressions for the mechanical steady-state
occupancy, we will now plot these functions for optimized parameter
values. We assume here that the classical cooperativity $C_{\text{cl}}$ is fixed (at
the maximal value permitted by the given experimental circumstances).
For purposes of demonstrating the benefit of varying $\theta$ in
the quantum regime of feedback cooling, we consider the limit of large
quantum cooperativity,
\begin{equation}
C_{\text{q}}\equiv C_{\text{cl}}/\bar{n}\gtrsim1.\label{eq:Large quantum cooperativity}
\end{equation}
This is the regime where the optical readout rate $\Gamma_{\text{meas}}$
of the mechanical position exceeds the thermal decoherence rate $\gamma_{\text{m}}\bar{n}$.
We focus here on the limit of ideal detection $\eta=1$.

In Fig.~\ref{fig:Func-gamma-fb} we plot $n_{\text{tot}}$ and its components, Eqs.~(\ref{eq:total number}-\ref{eq:correlation number}), as a function of the feedback parameter~$\sigma$.
We do so using the analytically optimized values $\theta_{\text{opt}}(\sigma)$,
Eq.~(\ref{eq:theta-opt}), and $\alpha_{\text{opt}}(\sigma)$
which is found by minimizing Eq.~(\ref{eq:solution for optimized theta}),
as the roots of higher order polynomials. This gives a sense of how
the achievable performance varies as the feedback gain is increased.
As expected from the discussion above, the ratio of thermal noise
to back action remains constant as $\sigma$ is varied
and is solely determined by the quantum cooperativity. For weak feedback
 $\sigma$ a large cut-off frequency can be afforded
(see Fig.~\ref{fig:Func-gamma-fb}, inset). Therefore an increase
in $\sigma$ leads to further suppression of thermal and
back-action noise. However, as $\sigma$ increases beyond
its optimal value, the influence of the imprecision noise must be
curbed by $\alpha\sim1$ for which no further suppression of the thermal
and back-action noise is possible without paying an even larger penalty
from the other sources. This trade-off determines the minimum achievable
value of $n_{\text{tot}}$. We note the following approximate scaling
$\alpha_{\text{opt}}\propto\sigma^{-1}$
from Fig.~\ref{fig:Func-gamma-fb} (inset), which explains, in view
of Eq.~(\ref{eq:solution for optimized theta}), why the ratio $n_{\text{fb}}/n_{\text{co}}$
is seen to be constant in Fig.~\ref{fig:Func-gamma-fb}. The
scaling also explains, in view of Eq.~(\ref{eq:theta-opt}), why the
optimal angle is approximately independed on the value of $\sigma$.

\begin{figure*}
\begin{centering}
\includegraphics[width=1\linewidth]{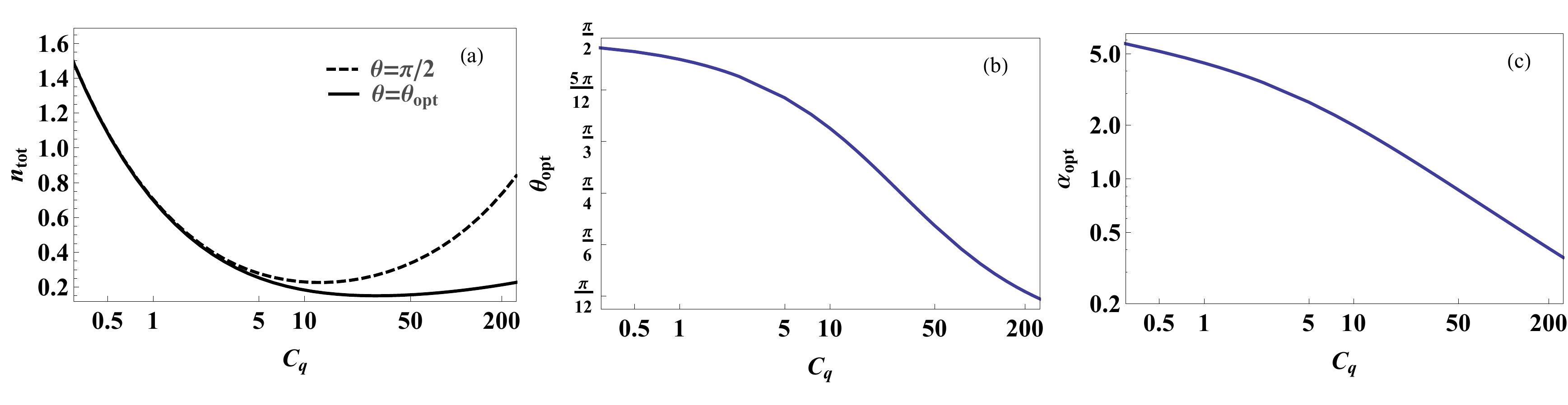}
\par\end{centering}
\caption{Optimized cooling for the same parameters as in Fig.~\ref{fig:Func-gamma-fb},
but with $C_{\text{q}}$ as an independent variable. (a) $n_{\text{tot}}$
as a function of $C_{\text{q}}$ evaluated at the optimal values $\theta_{\text{opt}}(C_{\text{q}}),\alpha_{\text{opt}}(C_{\text{q}}),\sigma(C_{\text{q}})$.
For comparison, we also plot the curve resulting from the suboptimal
choice $\theta=\pi/2$ evaluating at optimal parameter values given
this choice, $\alpha_{\text{opt}}^{(\pi/2)}(C_{\text{q}}),\sigma^{(\pi/2)}(C_{\text{q}})$.
Both functions have a local minimum, which can likely be ascribed
to the suboptimal functional form of the feedback gain $\mu(\omega)$
considered throughout this analysis. (b, c) Optimal
quadrature angle $\theta_{\text{opt}}(C_{\text{q}})$ and feedback
cut-off parameter $\alpha_{\text{opt}}(C_{\text{q}})$.\label{fig:ntot-func-Cq}}
\end{figure*}

Having determined the individual contributions to $n_{\text{tot}}$
and their behavior in the ``graphical minimization'' with respect
to the feedback strength $\sigma$, we plot the achievable
minimum occupancy $n_{\text{tot}}$ as a function of $C_{\text{q}}$
in Fig.~\ref{fig:ntot-func-Cq}. Subfigures a and b together clearly
demonstrate the necessity of adapting the measurement quadrature via
$\theta$ in order to get as close to the ground state as possible.
For small values of $C_{\text{q}}\lesssim1$ we find $\theta_{\text{opt}}\approx\pi/2$,
whereas $\theta_{\text{opt}}\rightarrow\pi/4$ as $C_{\text{q}}$
increases, which is consistent with the intuitive discussion in Sec.
\ref{sec:Heisenberg-microscope}. Note however that $n_{\text{tot}}(C_{\text{q}})$
in Fig.~\ref{fig:ntot-func-Cq}a exhibits a finite global minimum
for both the special case of $\theta=\pi/2$ and using the optimized
$\theta=\theta_{\text{opt}}(C_{\text{q}})$. For values of $C_{\text{q}}$
exceeding the minimum point, $\theta_{\text{opt}}$ drops below $\pi/4$
to compensate for the unnecessarily large (direct) back action. We
ascribe the appearance of the minimum in $n_{\text{tot}}(C_{\text{q}})$
to our suboptimal choice (\ref{eq:gain frequency}) for the feedback
gain function, $\mu(\omega)$. This can be understood in terms of suboptimal
mapping of the light quadratures to the mechanical mode as discussed
in Sec.~\ref{sec:Heisenberg-microscope}. Fig.~\ref{fig:ntot-func-Cq}c shows the optimal value of the feedback frequency cut-off
$\alpha_{\text{opt}}(C_{\text{q}})$, which approaches order unity
from above as $C_{\text{q}}$ increases towards the optimum. Similar
to the situation in Fig.~\ref{fig:Func-gamma-fb}, the decreasing
behavior of $\alpha_{\text{opt}}(C_{\text{q}})$ reflects the need
to suppress the bandwidth of the feedback noise when the feedback
gain increases.

From Fig.~\ref{fig:ntot-func-Cq} we conclude that it becomes relevant to choose a homodyne quadrature angle $\theta \neq \pi/2$ in the regime of very large quantum cooperativty, $C_{\rm q} \gg 1$. While the reduction in $n_{\rm tot}$ gained in this way will be small in absolute numbers, it can be significant relative to the value of $n_{\rm tot}$ achieved with $\theta=\pi/2$.

\section{Conclusion and outlook\label{sec:Conclusion}}

Simultaneous optimization of the feedback gain $\mu(\omega)$ and the
homodyne quadrature angle $\theta$ are crucial elements of realizing
the full potential of feedback cooling in the quantum regime.

As an outlook, we will point out two intriguing theoretical challenges
on the subject of feedback cooling in the quantum regime. The analysis
and optimization presented here was based on a particular choice for
the feedback filter function $\mu(\omega)$, as stated in Eq.~(\ref{eq:gain frequency}),
implementing a derivative-type feedback combined with a low-pass filter.
While this choice of $\mu(\omega)$ has desirable features for purposes
of cooling there is no reason to believe that the form of $\mu(\omega)$
assumed here is optimal from a theoretical perspective. Hence, this
work does not establish the ultimate limit of performance for the
scheme. This is an open question within quantum control theory (to
the knowledge of the authors) and merits further investigation. The authors of Ref.~\cite{WilsonSudhirPiroEtAl2015} made progress in this direction by deriving the optimal filter minimizing the position fluctuations for feedback based on measurements of the phase quadrature. As the authors emphasize,  this filter includes spectrally sharp features, which will require very fast electronics to implement and may not be feasible in practice. Accordingly, the
employed feedback filter was a delay line combined with a low-pass
filter \cite{WilsonSudhirPiroEtAl2015,SudhirWilsonSchillingEtAl2016}.
This choice is simpler to implement in practice, but is in fact mathematically
more cumbersome to treat than the form of $\mu(\omega)$ assumed here.

Another important remark should be made on the generality of the analysis
presented here. As mentioned in Sec.~\ref{sec:Model}, we adopted
here a  prescription for including feedback in the
Heisenberg-Langevin equations as done in several previous studies
found in the literature. While this approach can be justified rigorously for Markovian feedback, no such derivation is available for Non-Markovian feedback. Rigorous descriptions of Non-Markovian feedback exist based on the stochastic Schrödinger equation, cf.~\cite{Wiseman1994,Giovannetti1999}. However, it appears that a derivation of the corresponding Heisenberg-Langevin description has not been reported in the literature so far.  It is therefore possible that this and previous
studies neglect corrections due to the non-commutativity of Heisenberg operators at unequal times which may become significant in the quantum regime of operation. Only when these theoretical challenges have been addressed,
the ultimate quantum limits of feedback cooling can be established.

\section{References}
\bibliographystyle{unsrt}
\bibliography{FBCooling}

\appendix

\section{Stability condition\label{sec:stability}}

In this Appendix we derive the stability condition for the optomechanical
system subject to feedback. The validity of the linearized equations
of motion given in the main text hinges on the fulfillment of this
condition. Stability is ensured if the real part of all poles of the
effective mechanical susceptibility $\chi_{\text{eff}}(\omega)$ are
negative. The character of the poles can be determined using the Routh-Hurwitz
criterion \cite{DeJesus1987}, which in the present case leads to
a single non-trivial stability condition,

\begin{eqnarray}
1 & + & \sigma+\frac{1}{\alpha^{2}}+\frac{1}{Q_{\text{m}}\alpha}+\frac{\sigma}{Q_{\text{m}}\alpha} \\
  & + & \beta^{3}\biggl\{\frac{1}{Q_{\text{m}}}+\frac{1}{\alpha}+\alpha\biggr\} \nonumber \\
  & + & \beta^{2}\biggl\{1+\frac{1}{Q_{\text{m}}^{2}}+\frac{1}{\alpha^{2}}+\frac{2}{Q_{\text{m}}\alpha}+\frac{\alpha}{Q_{\text{m}}} \nonumber \\
  & + & \frac{\sigma}{Q_{\text{m}}}\left(\frac{1}{Q_{\text{m}}}+\frac{1}{\alpha}+\alpha\right)\biggr\} \nonumber \\
  & + & \beta\biggl\{\alpha+\alpha\sigma+\frac{2}{Q_{\text{m}}}+\frac{1}{Q_{\text{m}}\alpha^{2}}+\frac{1}{\alpha}+\frac{1}{Q_{\text{m}}^{2}\alpha} \nonumber \\
  & + & \frac{\sigma}{Q_{\text{m}}}\left(1+\frac{1}{Q_{\text{m}}\alpha}\right)\biggr\} \nonumber \\
  & - & \left(\sigma\beta^{2}+\frac{\beta\sigma}{\alpha}+\frac{\beta\sigma^{2}\alpha}{Q_{\text{m}}}\right) \nonumber > 0, \label{eq:stability-cond}
\end{eqnarray}
where we have introduced the optomechanical sideband resolution factor
$\beta\equiv 2\omega_{\text{m}}/\kappa$, the mechanical quality factor
$Q_{\text{m}}\equiv\omega_{\text{m}}/\gamma_{\text{m}}$, and (as
in the main text) the rescaled dimensionless feedback gain is $\sigma\equiv 2\mu_{\text{fb}}g_{\text{om}}\omega_{\text{m}}/\kappa\gamma_{\text{m}}$.
The feedback gain only enters the stability criterion (\ref{eq:stability-cond})
via the rescaled definition $\sigma$, which is why no explicit dependence
on the measurement quadrature angle $\theta$ enters. (If desired,
Eq.~(\ref{eq:stability-cond}) can be restated in terms of $\theta$
and the absolute gain $\tilde{\sigma}$ as pointed out in the main
text below Eq.~(\ref{eq:sigma-def}).) Note that in the main text we primarily focus on the idealized bad-cavity limit, $\kappa \rightarrow \infty$, in which the stability criterion~\ref{eq:stability-cond} is trivially fulfilled.

\section{Mechanical steady-state occupation $n_{\text{tot}}$ for arbitrary
$\kappa$\label{sec:Exact-solution-for}}

In the main text we focus on the idealized bad-cavity limit, $\kappa\rightarrow\infty$,
for simplicity of analysis. Here we will give the exact expression
for the mechanical steady-state occupation number, $n_{\text{tot}}$,
valid for arbitrary values of $\kappa$ (insofar as the stability
criterion of \ref{sec:stability} is fulfilled). Retracing
the steps taken in Sec.~\ref{sec:Steady-state-occupation} of the
main text, we decompose the mechanical occupancy into contributions
according to physical origin
\begin{equation}
n_{\text{tot}}=\frac{n_{X}+n_{P}}{2}-\frac{1}{2}=n_{\text{th}}+n_{\text{ba}}+n_{\text{fb}}+n_{\text{co}}+n_{v}-\frac{1}{2},\label{eq:C1}
\end{equation}
where $n_{X}$ and $n_{P}$ are the contributions from the position
and momentum variances, respectively. As will be clear from the expressions
below, the steady state of the mechanical mode in the presence of
feedback generally does not obey equipartition of energy, i.e., we
will have $n_{X}\neq n_{P}$. To display the position and momentum
contributions separately, we will decompose the various noise contributions
as $n_{i}=(n_{i,X}+n_{i,P})/2$ below. These will be expressed in
terms of the following dimensionless parameters: The classical optomechanical
cooperativity $C_{\text{cl}}=4g_{\text{om}}^{2}/\kappa\gamma_{\text{m}}$,
the rescaled feedback gain $\sigma=2\mu_{\text{fb}}g_{\text{om}}\omega_{\text{m}}/\kappa\gamma_{\text{m}}$,
the optomechanical sideband resolution parameter $\beta=2\omega_{\text{m}}/\kappa$,
the mechanical quality factor $Q_{\text{m}}=\omega_{\text{m}}/\gamma_{\text{m}}$,
and the feedback frequency cut-off parameter $\alpha\equiv\omega_{\text{fb}}/\omega_{\text{m}}$.
All expressions below are organized according to the parameters $\beta$
and $1/Q_{\text{m}}$, which are small in the typical parameter regime
of interest.

The exact value of the mechanical steady-state occupancy $n_{\text{tot}}$,
valid for all stable values of $\kappa$, is given by the following
contributions in Eq.~(\ref{eq:C1}): The thermal heating from intrinsic
mechanical damping,

\begin{equation}
n_{\text{th}}=\frac{n_{\text{th},X}+n_{\text{th},P}}{2}\label{eq:C2-2}
\end{equation}
\begin{eqnarray}
n_{\text{th},X} & = & \frac{1}{S_{\text{m}}}\left(\bar{n}+\frac{1}{2}\right)\Bigg[1+\frac{1}{\alpha^{2}}+\frac{1}{\alpha Q_{\text{m}}}\label{eq:C2}\\
 & + & \beta\left\{ \frac{1}{\alpha}+\alpha+\frac{1}{Q_{\text{m}}}\left(2+\frac{1}{\alpha^{2}}-\sigma\right)+\frac{1}{\alpha Q_{\text{m}}^{2}}\right\} \nonumber \\
 & + & \beta^{2}\left\{ 1+\frac{1}{\alpha^{2}}+\frac{1}{Q_{\text{m}}}\left(\frac{2}{\alpha}+\alpha+\frac{\sigma}{\alpha}\right)+\frac{1}{Q_{\text{m}}^{2}}\right\} \nonumber \\
 & + & \beta^{3}\left\{ 1+\frac{1}{\alpha^{2}}+\frac{1}{Q_{\text{m}}}\right\} \Bigg]\nonumber
\end{eqnarray}

\begin{eqnarray}
n_{\text{th},P} & = & \frac{1}{S_{\text{m}}}\left(\bar{n}+\frac{1}{2}\right)\Bigg[1+\frac{1}{\alpha^{2}}+\frac{1}{\alpha Q_{m}}\left(1+\sigma\right)\label{eq:C2-1}\\
 & + & \beta\left\{ \frac{1}{\alpha}+\alpha+\frac{1}{Q_{\text{m}}}\left(2+\frac{1}{\alpha^{2}}+\sigma\right)+\frac{1+\sigma}{\alpha Q_{\text{m}}^{2}}\right\} \nonumber \\
 & + & \beta^{2}\biggl\{ 1+\frac{1}{\alpha^{2}}+\frac{1}{Q_{\text{m}}}\left(\frac{2}{\alpha}+\alpha+\frac{\sigma}{\alpha}+\alpha\sigma\right) \nonumber \\
 & + &\frac{1+\sigma}{Q_{\text{m}}^{2}}\biggr\} \nonumber \\
 & + & \beta^{3}\left\{ 1+\frac{1}{\alpha^{2}}+\frac{1}{Q_{\text{m}}}\right\} \Bigg],\nonumber
\end{eqnarray}
the direct back-action heating,

\begin{equation}
n_{\text{ba}}=\frac{n_{\text{ba},X}+n_{\text{ba},P}}{2}\label{eq:C2-2-1}
\end{equation}

\begin{eqnarray}
n_{\text{ba},X} & = & \frac{1}{S_{\text{m}}}\frac{C_{\text{cl}}}{4}\Bigg[1+\frac{1}{\alpha^{2}}+\frac{1}{\alpha Q_{\text{m}}}\\
 & + & \beta\left\{ \frac{1}{\alpha}+\alpha+\frac{1}{Q_{\text{m}}}\left(2+\frac{1}{\alpha^{2}}-\sigma\right)+\frac{1}{\alpha Q_{\text{m}}^{2}}\right\} \nonumber \\
 & + & \beta^{2}\left\{ \frac{1}{Q_{\text{m}}}\left(\frac{1}{\alpha}+\alpha\right)+\frac{1}{Q_{\text{m}}^{2}}\right\} \}\nonumber
\end{eqnarray}

\begin{eqnarray}
n_{\text{ba},P} & = & \frac{1}{S_{\text{m}}}\frac{C_{\text{cl}}}{4}\Bigg[1+\frac{1}{\alpha^{2}}+\frac{1}{\alpha Q_{\text{m}}}\left(1+\sigma\right)\\
 & + & \beta\left\{ \frac{1}{\alpha}+\alpha+\frac{1}{Q_{\text{m}}}\right\} \Bigg],\nonumber
\end{eqnarray}
the imprecision noise heating,

\begin{equation}
n_{\text{fb}}=\frac{n_{\text{fb},X}+n_{\text{fb},P}}{2}\label{eq:C2-2-2}
\end{equation}

\begin{eqnarray}
n_{\text{fb},X}&= \frac{1}{S_{\text{m}}}\frac{\sigma^{2}}{2 C_{\text{cl}}}\Bigg[1+\beta\left\{ \alpha+\frac{1}{Q_{\text{m}}}\right\} \label{eq:C4} \\
&+ \beta^{2}\left\{ 1+\frac{\alpha}{Q_{\text{m}}}\left(1+\sigma\right)\right\} +\beta^{3}\alpha\Bigg]\csc^{2}(\theta),  \nonumber
\end{eqnarray}

\begin{eqnarray}
n_{\text{fb},P} &= \frac{1}{S_{\text{m}}}\frac{\sigma^{2}}{2 C_{\text{cl}}}\Bigg[1+\frac{\alpha}{Q_{\text{m}}}\left(1+\sigma\right)\label{eq:C4-1}\\
&+ \beta\left\{ \alpha+\frac{1}{Q_{\text{m}}}\left(1+\alpha^{2}+\alpha^{2}\sigma\right)+\frac{\alpha}{Q_{\text{m}}^{2}}\left(1+\sigma\right)\right\} \nonumber \\
&+ \beta^{2}\left\{ 1+\frac{\alpha}{Q_{\text{m}}}\left(2+\sigma\right)+\frac{\alpha^{2}}{Q_{\text{m}}^{2}}\left(1+\sigma\right)\right\} \nonumber  \\
&+ \beta^{3}\left\{ \alpha+\frac{\alpha^{2}}{Q_{\text{m}}}\right\} \Bigg]\csc^{2}(\theta). \nonumber
\end{eqnarray}
the correction term due to correlations between the direct back-action and
the measurement imprecision fluctuations ($n_{\text{co}}<0$ for parameter choices
of interest)

\begin{equation}
n_{\text{co}}=\frac{n_{\text{co},X}+n_{\text{co},P}}{2}\label{eq:C2-2-3}
\end{equation}

\begin{eqnarray}
n_{\text{co},X}&= -\frac{1}{S_{\text{m}}}\frac{\sigma}{4\alpha}\Biggl[1+\beta\left\{ 2\alpha+\frac{1}{Q_{\text{m}}}\right\} \label{eq:C5} \\
&+ \beta^{2}\left\{ \alpha^{2}+\frac{\alpha}{Q_{\text{m}}}\right\} \Biggr]\cot(\theta), \nonumber
\end{eqnarray}

\begin{eqnarray}
n_{\text{co},P}&= -\frac{1}{S_{\text{m}}}\frac{\sigma}{4\alpha}\Biggl[1+\frac{\alpha}{Q_{\text{m}}}\left(1+\sigma\right) \label{eq:C5-1} \\
&+ \beta\left\{ 2\alpha+\frac{\alpha^{2}}{Q_{\text{m}}}\left(1+\sigma\right)\right\} +\beta^{2}\alpha^{2}\Biggr]\cot(\theta), \nonumber
\end{eqnarray}
and finally the excess imprecision noise due to optical losses and
imperfect detection ($\eta<1$)

\begin{equation}
n_{v} =n_{\text{fb}}\left(\eta^{-1}-1\right),\label{eq:C2-2-4}
\end{equation}
where the denominator of these expressions is
\begin{eqnarray}
S_{\text{m}}&= 1 + \sigma+\frac{1}{ \alpha^{2}}+\frac{1}{\alpha \text{Q}_{\text{m}}}(1+\sigma)\label{eq:C7} \\
&+ \beta \biggl\{ \frac{1}{\alpha}\left(1-\sigma\right)+\left(\alpha+\frac{1}{\alpha \text{Q}_{\text{m}}^{2}}\right)(1+\sigma) \nonumber \\
&+ \frac{1}{\text{Q}_{\text{m}}}\left(2+\frac{1}{\alpha^{2}}+\sigma-\sigma^{2}\right) \biggr\} \nonumber \\
&+ \beta^{2} \biggl\{ 1+\frac{1}{\alpha^{2}}-\sigma+\frac{1}{Q_{\text{m}}}\left( \frac{1}{Q_{\text{m}}}+\alpha\right) \left( 1+\sigma\right) \nonumber \\
&+ \frac{1}{\alpha Q_{\text{m}}}\left(2+\sigma\right) \biggr\} \nonumber \\
&+ \beta^{3}\left\{ \frac{1}{\alpha}+\alpha+\frac{1}{Q_{\text{m}}}\right\} \nonumber .
\end{eqnarray}
These formulas generalize the expressions given in Ref.~\cite{Genes2008}
in which the special case $\theta=\pi/2$ was considered.

\section{Integrals over rational functions\label{sec:Rational-Integral}}

We consider integrals over rational functions of the
form
\[
\int_{-\infty}^{\infty}\frac{g_{n}\left(\omega\right)}{h_{n}\left(\omega\right)h_{n}\left(-\omega\right)}d\omega,
\]
where $g_{n}$ and $h_{n}$ are polynomials of the following form

\begin{equation}
g_{n}\left(\omega\right)=b_{0}\omega^{2n-2}+b_{1}\omega^{2n-4}+\cdots+b_{n-2}\omega^{2}+b_{n-1}
\end{equation}

\begin{equation}
h_{n}\left(\omega\right)=a_{0}\omega^{n}+a_{1}\omega^{n-1}+\cdots+a_{n-1}\omega+a_{n}
\end{equation}
and we assume that $a_{0}\neq0$ and that all the roots of $h_{n}\left(\omega\right)$
lie in the upper half plane. The solution to the integral can be stated in terms of the determinants of two square matrices $\Delta_{n},M_{n}$ of dimension $n$. Setting $a_{k}=0$ for any $k\notin\{0,\ldots,n\}$ the $(i,j)'\rm{th}$ entries of the matrices can be stated as
\begin{eqnarray}
(\Delta_{n})_{i,j}&=&a_{2j-i} \\
(M_{n})_{i,j} &=& \delta_{1,i}b_{j-1} + (1-\delta_{1,i})a_{2j-i},
\end{eqnarray}
where $\delta_{i,j}$ is the Kronecker delta function. The matrices differ only by their first rows as is clear from their explicit form,
\begin{equation}
\Delta_{n}=\left(\begin{array}{ccccc}
a_{1} & a_{3} & a_{5} & \ldots & 0\\
a_{0} & a_{2} & a_{4} & \, & 0\\
0 & a_{1} & a_{3} & \, & 0\\
\vdots & \, & \, & \ddots & \,\\
0 & 0 & 0 & \, & a_{n}
\end{array}\right),\,\,\,
\end{equation}

\begin{equation}
M_{n}=\left(\begin{array}{ccccc}
b_{0} & b_{1} & b_{2} & \ldots & b_{n-1}\\
a_{0} & a_{2} & a_{4} & \, & 0\\
0 & a_{1} & a_{3} & \, & 0\\
\vdots & \, & \, & \ddots & \,\\
0 & 0 & 0 & \, & a_{n}
\end{array}\right).
\end{equation}
The value of the integral can now be stated as

\begin{equation}
\int_{-\infty}^{\infty}\frac{g_{n}\left(\omega\right)}{h_{n}\left(\omega\right)h_{n}\left(-\omega\right)}d\omega=i(-1)^{n+1}\frac{\pi}{a_{0}}\frac{\det M_{n}}{\det \Delta_{n}}.\label{eq:Integral Solution}
\end{equation}
This formula was used to evaluate $n_{\text{tot}}$ in terms of integrals
over spectral densities as presented in the main text. The formula~(\ref{eq:Integral Solution}) appears in Ref.~\cite{GRADSHTEYN1980211} on Page 253, however, there the factor $(-1)^{n+1}$ was omitted.
\end{document}